\documentclass[useAMS,usenatbib]{mnras}

\usepackage{url,times,graphicx,amsmath,amsfonts,amssymb,aas_macros,color,epsfig,epstopdf}

\newcommand{\mhalo}{M$_{\rm halo}$}
\newcommand{\mstar}{M$_{\rm star}$}
\newcommand{\lcdm}{$\Lambda$CDM}
\newcommand{\msun}{M$_\odot$}
\newcommand{\vlos}{V$_{\rm los}$}
\newcommand{\vmax}{V$_{\rm max}$}
\newcommand{\vlosMs}{V$_{\rm los}-$M$_{\rm star}$}

\newcommand{\Vmax}{$V_{\rm max}$}

\def\kms{km$\,$s$^{-1}$}

\newcommand{\hMpc}{{\ifmmode{h^{-1}{\rm Mpc}}\else{$h^{-1}$Mpc}\fi}}
\newcommand{\hkpc}{{\ifmmode{h^{-1}{\rm kpc}}\else{$h^{-1}$kpc}\fi}}
\newcommand{\hMsun}{{\ifmmode{h^{-1}{\rm {M_{\odot}}}}\else{$h^{-1}{\rm{M_{\odot}}}$}\fi}}
\newcommand{\ltsima}{$\; \buildrel < \over \sim \;$}
\newcommand{\gtsima}{$\; \buildrel > \over \sim \;$}
\newcommand{\lsim}{\lower.5ex\hbox{\ltsima}}
\newcommand{\gsim}{\lower.5ex\hbox{\gtsima}}
\def\LCDM{$\Lambda$CDM}

\def\lesssim{\mathrel{\hbox{\rlap{\hbox{\lower4pt\hbox{$\sim$}}}\hbox{$<$}}}}
\def\gtrsim{\mathrel{\hbox{\rlap{\hbox{\lower4pt\hbox{$\sim$}}}\hbox{$>$}}}}

\newcommand{\beq}{\begin{equation}}
\newcommand{\eeq}{\end{equation}}
\def\beqa{\begin{eqnarray}}
\def\eeqa{\end{eqnarray}}
\def\hMpc{$h^{-1}\,{\rm Mpc}$}
\def\hkpc{$h^{-1}\,{\rm kpc}$}
\def\LCDM{\ensuremath{\Lambda}CDM}

\def\Vmax{$V_{\rm max}$}

\def\head{
 \vbox to 0pt{\vss
                   \hbox to 0pt{\hskip 440pt\rm LA-UR-10-07069\hss}
                  \vskip 25pt}}

\title[Signatures of Dark Matter Halo Expansion]
{Signatures of Dark Matter Halo Expansion in Galaxy Populations}
\author[Brook \& Di Cintio]
       {Chris B. Brook$^{1}$\thanks{E-mail: cbabrook@gmail.com} \& Arianna Di Cintio$^{2}$\\
$^{1}$Ramon y Cajal Fellow, Departamento de F\'isica Te\'orica, Universidad Aut\'onoma de Madrid, 28049 Cantoblanco, Madrid, Spain\\
$^{2}$Dark Fellow, Dark Cosmology Centre, NBI, University of Copenhagen, Juliane Maries Vej 30, DK-2100 Copenhagen, Denmark\\}

\begin{document}

\date{Accepted XXXX . Received XXXX; in original form XXXX}

\pagerange{\pageref{firstpage}--\pageref{lastpage}} \pubyear{2010}

\maketitle

\label{firstpage}


\begin{abstract}

Dark matter cores within galaxy haloes can be formed by energy feedback from star forming regions: an energy balance suggests that the maximum \textit{core formation efficiency} arises in galaxies with \mstar$\sim10^{8.5}$\msun.
We show that a model population of galaxies, in which the density profile has been modified by such baryonic feedback, is able to explain the observed galaxy velocity function and Tully-Fisher relations significantly better than a model in which a universal cuspy density profile is assumed.

Alternative models, namely warm or self-interacting dark matter, also provide a better match to  these observed relations than a universal  profile model does, but make different predictions for how halo density profiles vary with mass compared to the baryonic feedback case. We propose that different core formation mechanisms may be distinguished based on the imprint they leave on galaxy populations over a wide range of mass.

Within the current observational data we find evidence of the expected signatures of the mass dependence of core formation generated by baryonic feedback.

\end{abstract}

\noindent
\begin{keywords}
 galaxies: evolution - formation - haloes cosmology: theory - dark matter
 \end{keywords}

\section{Introduction} \label{sec:introduction}

The rotational velocity of gas and stars  in galaxy discs implies that galaxies are embedded within dark matter `haloes'.
A central tenet of the canonical cold dark matter (\lcdm) cosmological model  is that such haloes have dense  central regions with  steep, ``cuspy" inner density profiles (\citealt{navarro96}, hereafter NFW profile). However,  observed  rotation curves imply that dwarf galaxies have flat ``cores" at their centre \citep{moore94}.   This ``cusp-core" crisis is a major  challenge  to the \lcdm\ paradigm,  heightened by studies which demonstrate that some non-standard dark matter particles form cored haloes \citep{maccio12,vogelsberger12}.

On the other hand, astrophysical processes \citep{navarro96b,pontzen12} within a cold dark matter framework may modify halo density profiles and create cored haloes at specific halo masses \citep{DiCintio2014a}.  While observed rotation curves of individual galaxies provide compelling evidence for the existence of cores in low surface brightness and dwarf galaxies \citep{oh11}, they are not able to distinguish between the various proposed origins of cores.
Therefore, it is essential to establish whether the ``cusp-core"  crisis is pointing to a new cosmological paradigm, or is instead  the result of astrophysical processes within the standard \lcdm\ model.

In the \lcdm\ paradigm, cores may result from the non-adiabatic impact of gas outflows on dark matter haloes. There is significant evidence \citep{weiner09,martin12} that energy feedback from  star-formation activity  drives gas out of galaxies, and that processes such as
 radiation energy from massive stars,  stellar winds and supernova explosions  are important  in  galaxy formation \citep[e.g.][]{brook11}. The degree to which such energetic processes  flatten the inner density profiles of haloes depends on the ratio of their stellar to halo masses \citep[][see Section~2 for an analytic argument]{DiCintio2014a}.  
 
 Low mass galaxies are dark matter dominated: those with \mstar$\lsim$3$\times$10$^6$M$_\odot$ do not produce enough energy to flatten their halo's  inner density profile, which remains steep \citep{penarrubia12,governato12}.  As stellar mass increases relative to the dark matter mass,  the inner density profile becomes increasingly flat \citep{governato12}.  The flattening is greatest when   \mstar$\sim$3$\times$10$^8$\msun, after which the increasingly deep  potential well is able to  oppose halo expansion   resulting in less  flattening for more massive galaxies \citep{DiCintio2014a}.

One approach to test core formation by baryonic processes has been to study low mass dwarf galaxies, where the theory predicts cuspy, NFW profiles. A multitude of studies have attempted to determine whether galaxies such as Sculptor have a core \citep{Battaglia08,walker09,evans09,walker11,agnello12,amorisco12,laporte13,adams14,strigari14}, without a clear  consensus emerging.

In this study, we therefore take another approach and explore the properties of an ensemble of galaxies.
We  show that the mass dependence of cores formed through baryonic processes should leave an imprint on galaxy populations, and that such imprint  may be used to  distinguish between different  core formation mechanisms.
Indeed, while alternative dark matter particles could also be able to create cores, they would result in a different dependence of core sizes versus galaxy mass.

In particular, we study the velocity function of galaxies \citep{zavala09,zwaan10,tg11,papastergis11,obreschkow13,klypin14}, as well as the relation between velocity and stellar mass \citep{tully77}, over a wide mass range in which core sizes are expected to vary in different ways according to different models.

To model haloes that are flattened by energetic feedback processes, we use the  mass dependent  density profile \citep{DiCintio2014b}, hereafter DC14 profile, in which   galaxies with \mstar\ between $\sim$3$\times$$10^6$ and $\sim$$10^{10}$ \msun\ have central densities shallower than the NFW profile \citep{DiCintio2014a}. 
The DC14 model has already been used to study the variation in core sizes for Local Group dwarf galaxies, in which their observed kinematics has been explained in terms of a density profile dependent on \mstar/\mhalo\  \citep{brook15}.

Standard alternative dark matter models, in particular a warm dark matter (WDM) and a self interacting dark matter (SIDM) one, are also explored and the expected properties of galaxy populations within such models are compared with the fiducial NFW model. We highlight the manner in which the alternative dark matter models presented can be generalized to a wide range of parameterizations.

In Section~2 we present the ingredients of our models, including an analytic argument for our expectation of how core sizes vary with mass in the baryonic feedback scenario.
In Section~3 we create velocity distribution functions of galaxies and \mstar-velocity relations, and we compare the predictions coming from the DC14 model to the results obtained by
using a standard NFW model (Sections~3.1 and ~3.2). We also compare velocity functions and \mstar-velocity relations for the alternative dark matter models, namely WDM and SIDM (Section~3.3).
We discuss possible uncertainties and caveats in Section~3.4 and ~3.5.
In Section~4 we show that the different models result in different observational trends with mass within galaxy populations, and that such trends  may be distinguishable with careful observations.


\begin{figure}
\hspace{-.25cm}
\includegraphics[width=.5\textwidth]{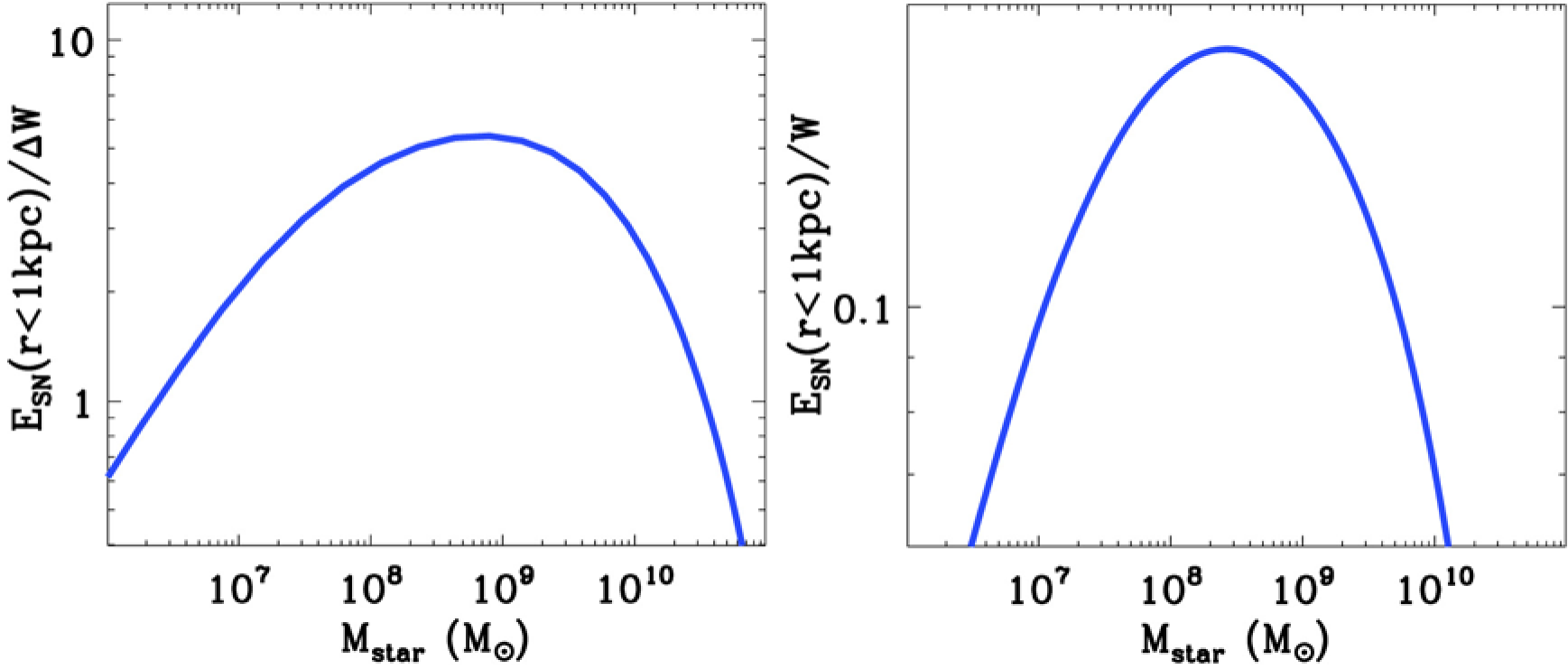}
\caption{
Left Panel:The ratio of supernova energy from stars within the central 1$\,$kpc, E$_{SN}$(r$<$1$\,$kpc), and the energy required to transform an NFW profile to a profile with a 1$\,$kpc core ($\Delta$W),
 plotted as a function of \mstar.  Only galaxies in the mass range, $3\times$10$^6$$\lsim$ \mstar$\lsim$10$^{10}$\msun\ have sufficient energy to form cores. Right Panel:   
  The ratio of supernova energy from stars within the central 1$\,$kpc, and  the potential of an NFW profile with embedded baryonic disk}  as a function of \mstar. This ratio gives an indication of the efficiency of core formation: one  expects the largest cores at \mstar$\sim$3$\times$10$^8$\msun.

\label{fig:analytic}
\end{figure}

\section{\bf Ingredients of our models}
Our models consist of a mass function for dark matter haloes with appropriate concentrations, a relation between stellar and halo mass, and exponential stellar and gas discs.  Dark matter haloes are modelled by different density profiles, as explained below. Halo mass distributions for \lcdm\ and WDM assume Planck cosmological parameters and are created with {\tt HMFcalc} \citep{HMFcalc13}, with appropriate corrections to the WDM mass function \citep{schneider12}. The  WDM model has a particle mass of 2$\,$KeV, with the effect of using 1, 2 \& 3$\,$KeV particles  shown in section~\ref{wdmvary}.

We use  the concentration-mass relation from Planck \lcdm\ \citep{Dutton14}, and adjust the  WDM halo concentration \citep{schneider12}. An adiabatic contraction correction to the dark matter haloes is applied to all models, as found in cosmological simulations for galaxies as massive as the Milky Way \citep{DiCintio2014b}. Such contraction alters the concentration (i.e. the scale radius) rather than the inner slope, allowing core sizes to be set. Specifically we use
\begin{equation}
{\rm C}=(1.0+ 0.00001e^{3.4{\rm X}})\times {\rm C}_{\rm NFW}\\
 \label{c}
\end{equation}
where X$=\log_{10}$(\mstar/\mhalo) + 4.5, and C$_{\rm NFW}$ is the concentration given from the adopted  \citep{Dutton14} relation.\\
\vspace{-.35cm}

Haloes are matched to galaxy stellar masses using the empirical abundance matching relation of \citet{guo10}. We explore the effects of different abundance matching relations in Section~\ref{section:ab}.

Once a relation between dark matter halo mass and galaxy stellar mass has been set, we add a stellar exponential disc with scale-length h$_{s}$ taken from the observed \mstar-h$_s$ relation, log$_{10}$h$_{\rm s}$=$-$2.462+0.281log$_{10}$\mstar (equation 7 in \citet{dutton11b}).
We further add a gas disk  with  mass M$_{\rm gas}$$=$1.3M$_{\rm HI}$, with log$_{\rm 10}$(M$_{\rm HI}$/\mstar)$=$$-0.43$log$_{10}$\mstar$+3.75$  \citep{papastergis12} and  disc scale length h$_{g}$=3h$_{s}$.

The circular velocity of a thin exponential disk of mass M$_{\rm d}$ and scale length h$_{\rm i}$ is given by:

\begin{equation}
\rm{V}_{disk}^2(r)=\frac{GM_d}{h_i}2y^2[I_0(y)K_0(y) - I_1(y)K_1(y)]
\end{equation}

\noindent where $y=r/(2h_i)$ and $I_n,K_n$ are modified Bessel functions \citep{dutton11}.

In all models we include the universal dark matter fraction of mass ($\Omega_{\rm DM}$/$\Omega_{m}$=0.85) in dark matter haloes. 
The density profile of dark matter haloes is set to be either the universal, NFW one \citep{navarro96}, or the  mass-dependent DC14 density profile described in the next section \citep{DiCintio2014b}.

WDM halos are known to have cored profiles \citep{tremaine79,bode01}. However,  \cite{maccio12} find that WDM particles with the masses  considered in our study, 1-3KeV, result in very small cores  of order 10pc for halos of mass 10$^{10}$\msun, and that it requires much lower mass ($\sim$0.1KeV) WDM particles to create  cores of 1\,kpc. In this study, we approximate the very small cores of  WDM halos in our model by retaining the NFW profile.

 In the adopted SIDM model,  core sizes are set to zero for Milky Way mass galaxies  (\mstar=6$\times$10$^{10}$\msun) and increases as the scale-length decreases, until reaching a core size of 1$\,$kpc for \mstar$=$10$^8$\msun\ \citep{vogelsberger12,vogelsberger14}. Lower mass  haloes all have core sizes of 1$\,$kpc. 
Other parameterizations for varying core sizes are explored in section~\ref{corevary}.  Of course, they do not cover all SIDM models and it is possible to fine tune SIDM parameters in order to have a mass variation of halo profile similar to the one of baryonic outflow models. On the other hand, the parameterisations that we do explore here can be generalized, since they are representative of a range of alternative dark matter candidates that predict similar behaviour of core size with galaxy mass. 

\noindent Cored profiles of dark matter for the SIDM case take the form: 

\begin{equation}
 \rho(r)=\frac{\rho_0 r_{scale}^3}{(r_c+r)(r_{scale}+r)^2}
 \end{equation}
 
\noindent where $\rho_0$ is a characteristic halo density, $r_{scale}$ is a scale radius and $r_c$ is the core size \citep{penarrubia12}.  
The velocity profile of dark matter haloes,  $\rm{V}_{DM}$, will thus depend on the type of density profile chosen.
The theoretical, total circular velocity profile of a galaxy at any radius, defined as V$_{\rm rot}$, is therefore given by the quadratic sum of the circular velocities of the various components:

\begin{equation}
 \rm{V}_{rot}(r)=\sqrt{\rm{V}_{star}^2(r) +  \rm{V}_{gas}^2(r) +  \rm{V}_{DM}^2(r)}
\end{equation}

 \subsection{The mass dependent `DC14' profile}\label{DC14}
 The analytic work in this section follows the work of \cite{penarrubia12} who model the balance between the gravitational potential of galaxies  and the energy from central star formation regions as the main determinant of core formation in the case of cores created by astrophysical processes \citep{penarrubia12}.  Once an exponential stellar and gas discs have been set as explained above, using observed scaling relations \citep{dutton11b,fathi10,papastergis12}, it is possible to compute the supernova energy from stars within the central 1$\,$kpc of each galaxy and compare with the depth of its potential well.
 
The left panel  of  Figure~\ref{fig:analytic}  shows the ratio of such supernova energy, E$_{SN}$(r$<$1$\,$kpc), divided by the gravitational potential energy $\Delta$W required to transform an initially NFW halo to a halo whose profile has a 1$\,$kpc core, plotted as a function of \mstar.

\noindent$\Delta$W is defined as $\Delta$W=(W$_{cusp}$-W$_{core}$)/2, where:
\begin{equation}
W=-4\pi G\int_0^{rvir}\rho(r)M(r)rdr
\end{equation}
\noindent The supernova energy within 1 Kpc is easily computed by:
\begin{equation}
E_{SN}=\rm M_{\rm star}(<1kpc)\bar{f}_{SN}\epsilon10^{51}erg
\end{equation}

\noindent where $\bar{f}_{SN}$ is the fraction of stars that will explode as supernovae divided by the mean stellar mass according to the initial mass function selected ($\bar{f}_{SN}=0.00925$ for a \citet{kroupa02} IMF), $\epsilon$ is the fraction of energy that will be transferred into the dark matter halo and $10^{51}$erg is the energy released during a single supernova event. 

Figure \ref{fig:analytic} assumes that  galaxies are connected to  dark matter haloes via the abundance matching relation of \citep{guo10}. It is clear that only galaxies in the mass range 3$\times$10$^6$$\lsim$ \mstar$\lsim$$\times$10$^{10}$\msun\ have sufficient energy to form cores.

The right panel of Figure~\ref{fig:analytic} shows the ratio of supernova energy from stars within the central 1$\,$kpc, and  the gravitational potential energy of an NFW halo  plus a baryonic disk as a function of \mstar. This ratio, namely the balance between SNe energy vs potential of the galaxy, gives an indication of the efficiency of core formation: the largest cores are expected at the peak of this relation, i.e. at  \mstar$\sim$3$\times$10$^8$\msun. 
 
 In Figure~\ref{fig:analytic} we used $\epsilon=0.4$, and we note that using different values will simply shift the plots up or down, changing the minimum mass for core formation but without modifying the position of the peak. 
 Whilst we have included only energy from SNe in this calculation, it is likely that ionising radiation will help drive outflows \citep{murray10}: again our analytic argument of where we expect the peak of core formation will not be affected, as long as this feedback scales with star formation rate.

We will explore the effects of AGN, which do not scale  with star formation rate, later in the paper; for now we just note that the stellar mass where core formation is most efficient is below the mass range where we expect AGN feedback to be important. 
Bulges have not been included in our models because at the core formation peak mass in Figure~\ref{fig:analytic} most galaxies do not have bulges. Nevertheless, we tested the results of adding empirically motived bulges, following the Hernquist formalism \citep{hernquist90}: we found that the only difference is the appearance of  a ``shoulder" on the right side of the the plots, since more SNe energy is stored within the inner 1\,kpc of a galaxy with a bulge, while the peak of core formation does not get affected.
  
 Of course, our model is very simplistic. In particular, \cite{penarrubia12} showed that the energy requirements for core formation  are significantly lower at higher redshift. And further, \cite{maxwell15} recently showed that the radius to which dark matter is moved when it is expelled from the cusp region, will also have an effect on the energy required to form a core.
Yet a very similar relation between core formation and galaxy mass has been found in cosmological simulations  \citep{DiCintio2014a} of galaxies that match a wide range of galaxy scaling relations \citep{brook12b,stinson13,kannan14,obreja14}: within these simulations,  the flattest profiles are also found at  \mstar$\sim$3$\times$10$^{8}$\msun.

The mass dependent DC14 density profile is  based on such cosmological simulations, and accounts for the expansion of dark matter haloes due to  the effects of feedback from star forming regions  \citep{DiCintio2014b}. These profiles  self consistently account for the distance to which dark matter is moved, addressing the concerns of \cite{maxwell15}, and  take the form:
\begin{equation}
\rho(r)=\frac{\rho_s}{\left(\frac{r}{r_s}\right)^{\gamma}\left[1 + \left(\frac{r}{r_s}\right)^{\alpha}\right] ^{(\beta-\gamma)/\alpha}}
\label{five}
\end{equation}
\noindent where ($\alpha,\beta,\gamma$) indicate the sharpness of the transition, the outer and the inner slope, respectively \citep{merritt06}.

The DC14 model describes profiles that have a range of inner slopes, with the parameters of the double power law model being dependent on the stellar-to-halo mass ratio  of a galaxy in the following way \citep{DiCintio2014b}:

\begin{equation}
\begin{aligned}
&\alpha= 2.94 - \log_{10}[(10^{X+2.33})^{-1.08}  +  (10^{X+2.33})^{2.29}]\\
&\beta=4.23+1.34X+0.26X^2\\
&\gamma= -0.06 + \log_{10}[(10^{X+2.56})^{-0.68}  +  (10^{X+2.56})]
\end{aligned}
\label{abg}
\end{equation}

\noindent where $X=\log_{10}($\mstar/\mhalo).
The scale radius, $r_s$, is connected to the concentration of the halo, which varies with mass.

As long as the flattening of the central dark matter density is mass dependent, and there is a maximum flattening at a similar stellar mass as found in both the analytic model and the simulations, the exact form of the profile will not affect the  arguments presented in this study.

\begin{figure*}
\includegraphics[width=1.\textwidth]{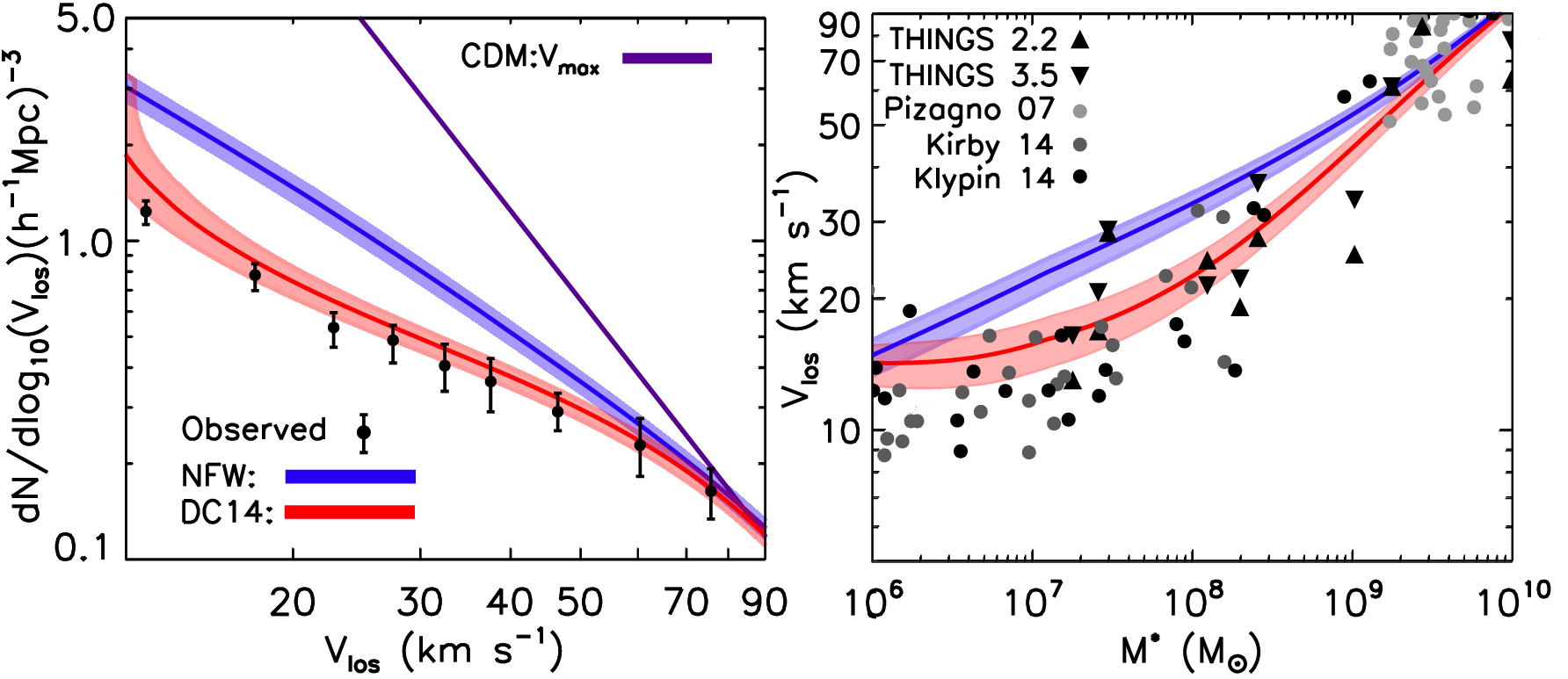}
\caption{Left Panel: the number of galaxies per line of sight velocity per unit volume, as a function of  maximum circular velocity (\vmax) for CDM theory (purple line) and as a function of \vlos\ for observations. The observed  distribution of \vlos\ is shown as black circles with error bars \citep{klypin14}. The \vlos\ distribution for a \lcdm\ model with NFW and DC14 profiles are shown in blue and red, respectively, with the velocity measured at the radius indicated in Figure~\ref{rmeasure}. Shaded regions show the same distribution using $\pm$20\% the radius at which measurements are taken.The mass dependent DC14 profile results in a  flattening of the \vlos\ distribution, caused by increasingly flat density profiles as we move from disc galaxies with high \vlos\ to lower \vlos. At the lowest, dwarf scales, the density profiles of the DC14 model steepen again, reflected in an upturn at \vlos\ $ \lsim $ 20$\,$km$\,$s$^{-1}$, in agreement with observation. Right Panel: The observed (filled symbols) and predicted \mstar-\vlos\ relations for the \lcdm\ model with NFW  (blue) and DC14 (red) profiles, measured at 2.2 and 3.5 disc scale-lengths with solid line at the midpoint, 2.85 disc scale-lengths.  For the  DC14 profile, the relation bends down from the standard NFW model at intermediate masses, and then flattens at low stellar masses, providing a match for observed  dwarf galaxies.} 
\label{fig:CDM}
\end{figure*}

\section{Results}
\subsection{Velocity Function}
When testing theoretical cosmological models,  dark matter haloes can be matched  to  observed galaxies  within a given volume \citep[e.g.][]{guo10,moster10,papastergis14}.  The density profiles of  dark matter haloes, together with the above mentioned disk distributions of stars and gas,  can be used to determine the  theoretical velocity profile of a galaxy at a given radius, $\rm{V}_{rot}$, according to Eq.3: such velocity is determined by the total mass of dark matter, gas and stars contained within that radius.

The standard \lcdm\ model, in which haloes have cuspy, NFW density profiles \citep{navarro96}, predicts a far greater number of galaxies than are observed with line-of-sight velocity (\vlos) less than  $\sim$60 km$\,$s$^{-1}$ \citep{zavala09,zwaan10,tg11,papastergis11,klypin14}.

This can be seen by comparing the \lcdm\ prediction, purple line, to the observational data points \citep{klypin14} in the left panel of Figure~\ref{fig:CDM}. The observed  \vlos\ distributions  are taken from compiled data of the of HI  line-widths of Local Volume galaxies \citep{karachentsev13} measured at 50\% of the maximum intensity, $W50$, with a further accounting for the number of early type galaxies which do not have significant amounts of HI \citep{klypin14}.

The  comparison is made assuming that the observed velocities of gas within galaxies track the maximum possible velocity associated with the corresponding dark matter halo, \vmax.  
The assumption made in \citet{klypin14} that observed velocities relate directly to \vmax\ may be justified in high mass galaxies, but is not necessarily correct in less massive galaxies, where  the rotation curves  are often still increasing at their outermost measured point \citep{catinella06,swaters09,deblok08,oh11}. For example, in a dwarf  galaxy with stellar mass \mstar=$10^{7-8}$\msun,   observed velocity measurements often come from stars within 1-2$\,$kpc of the galaxy centre, while the \vmax\ of the corresponding NFW halo  is at $\sim$7-10$\,$kpc. If haloes are  cored or expanded, then the \vmax\ may be even  further  from the centre of the galaxy: indeed, the maximum observed rotation velocity may be a poor estimate of \vmax\ in the case of a cored profile.

In \citet{papastergis14} this issue is addressed by attempting to place observed galaxies on the V$_{\rm rot}$-\vmax\ diagram expected from abundance matching, by fitting several dark matter haloes to observed rotation curves and considering only the last measured point of such velocity curves in order to determine the maximum velocity of the most massive halo that fits such point (see also \citet{ferrero12}).  In this sense, this is a more conservative approach than the one followed by \citet{klypin14}. However such methodology still assumes that dark matter haloes follow the universal NFW profile, but for some of the observed rotation curves in low mass galaxies this is not the case (see for example UGC7577 in their Fig.5).

Debating whether rotation curves {\it appear flat} may not be the optimal manner to address the issue of how to best compare theory and observation. Models provide predictions for velocities at all radii, so one can  compare observations with  theoretical predictions for velocities at any radius, as long as the two are done consistently. This is particularly important when comparing models of different density profiles. 

We therefore suggest a different approach, which is to estimate the physical radius at which measurements are made, and then compare models and observations at that same radius. 
In order to determine the radial region where \vlos\ is measured in HI surveys, we  find the radius, R$_{\rm m}$, at which observed rotation curves reach the inclination and velocity dispersion adjusted value V$_{\rm rot}$.

\begin{figure}
\includegraphics[width=.45\textwidth]{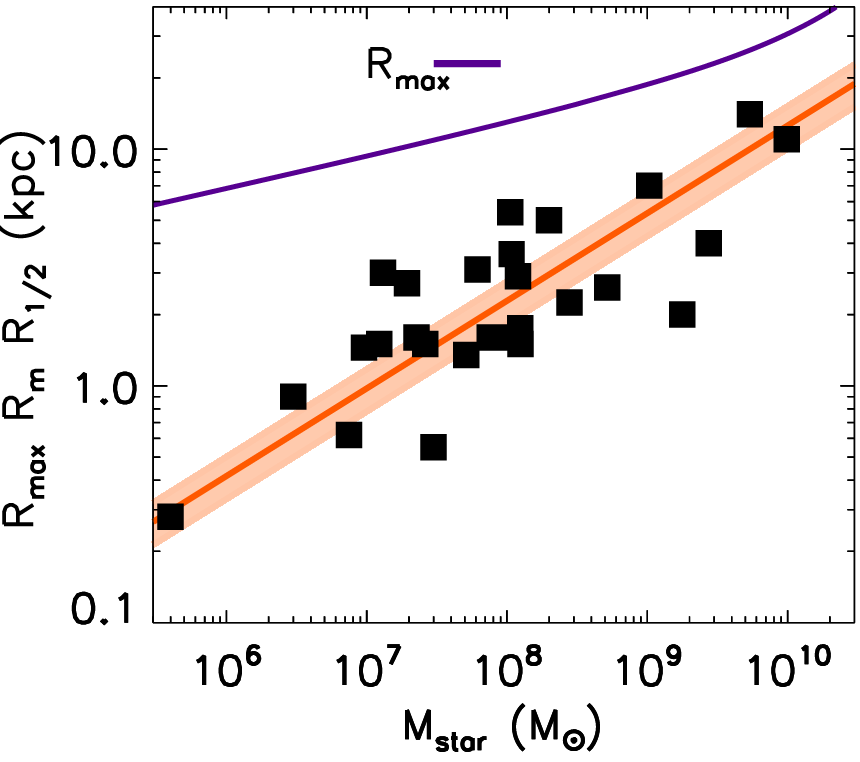}
\caption{The radius, R$_{\rm m}$, at which the observed rotation curves  of 26 observed galaxies taken from the literature  reach the  inclination and velocity dispersion adjusted value V$_{\rm rot}$$=$$\sqrt{((W50/2/sin(i))^2-\sigma_v^2)}$, plotted against stellar mass (black squares). The orange line is a fit to this data, with the shaded region indicating R$_{\rm m}$$\pm$20\%.
The purple line is the radius R$_{\rm max}$ at which an halo of a specific stellar mass \citep{guo10} reaches its maximum velocity \Vmax. The discrepancy between R$_{\rm max}$ and R$_{\rm m}$ is more severe moving toward low mass galaxies.}
\label{rmeasure}
\end{figure}

In Figure~\ref{rmeasure} we plot, as black squares, R$_{\rm m}$ as a function of \mstar\ for 26 observed  galaxies which have \mstar$\lsim$10$^{10}$\msun, high resolution rotation curves \citep{oh08,deblok08,oh11,cannon11,oh15} and corresponding $W50$ values \citep{walter08}. The orange line shows a fit to the data,
\begin{equation}\label{rm}
\rm log(R_{\rm m})=-2.75+0.37\times \rm log(M_{\rm star}),
\end{equation}  
with the shaded region indicating R$_{\rm m}$$\pm$20\%.
The purple line is the radius R$_{\rm max}$ at which a halo of a specific stellar mass, according to \mstar-\mhalo\ relations, reaches its maximum velocity \Vmax.
It is clear how moving toward low mass galaxies the gas is tracking an increasingly inner region of the halo rather than extending to R$_{\rm max}$.
The lower is the mass of a galaxy, the higher is the separation between R$_{\rm m}$ and R$_{\rm max}$.
We therefore decided to use the radius R$_{\rm m}$ as the radius where we measure velocities in our theoretical models, to be consistent with observations.

In the left panel of Figure~\ref{fig:CDM}, we derived theoretical velocity functions by first assuming a mass function for dark matter haloes, with appropriate halo concentration-mass dependence,  then using the density profiles of different models to derive galaxies'  circular velocities at each radius. We then measure   circular velocities at the radius  $\rm{R_m}$ to mimic the region where observations are made. We remark that before comparing the observed line-of-sight velocity, \vlos, with the theoretically predicted rotation curve, $\rm{V}_{rot}$, a correction must be applied to account for the pressure support of gas in real galaxies.
 When using HI line widths, \vlos$=$$W50/2$, where $W50$ is the width of the HI line profile measured at 50 per cent of the peak value, and the rotational velocity is then:
\begin{equation}
 \rm{V}_{rot}=\sqrt{\left(\frac{\rm V_{los}}{sin(i)}\right)^2-\sigma_v^2},
 \label{eq:inc}
\end{equation}
 where  $i$ is the inclination of the galaxy and $\sigma_v$=8 km/s is the typical amplitude of gas turbulent motion, in addition to rotation \citep{papastergis14}.
 In our theoretical velocity function, we average over all possible inclinations.    We use equation~\ref{eq:inc} to transform from circular velocities measured at R$_{\rm m}$ to \vlos\ velocities, and do this self-consistently for all the models, i.e. NFW, DC14, WDM and SIDM.

Using the standard  \lcdm\ model with  associated cuspy, NFW halo density profiles, the model still overpredicts the number of galaxies with  \vlos$\lsim$60$\,$km$\,$s$^{-1}$: the theoretical NFW velocity function  lies  significantly above observations as  can be seen from the blue shaded line in Figure~\ref{fig:CDM}.   

 This discrepancy motivates the exploration of other models which may lead to different   predictions for  the distribution of  \vlos\  of galaxies.
We first explore the mass dependent DC14 density profile, based on simulations where cores are formed through outflows driven by star formation activity, and fully described in Section~\ref{DC14}.

The theoretical velocity function based on a DC14 profile is shown as a red shaded region in Figure~\ref{fig:CDM} and its agreement with observations is notable.
 As we  move from massive to intermediate mass galaxies, the DC14 profile describes haloes which become increasingly flat in their inner regions, resulting in fewer galaxies with 30$\,\lsim\,$\vlos$\,\lsim\,$60$\,$km$\,$s$^{-1}$, in line with observations.
Below \mstar$\sim$3$\times$10$^8$\msun, the inner region of   DC14 profiles become  increasingly steep. This results in a slight upward bend at  \vlos$\lsim$20$\,$km$s^{-1}$ with the \vlos\ distribution eventually joining the standard \lcdm\ model for \vlos$\lsim$10$\,$km$s^{-1}$, where the haloes retain a  NFW profile given the insufficient energy from SNe to create a core.

This upward trend in the velocity function at  \vlos$\lsim$20$\,$km$s^{-1}$ is a prediction of  models in which cores are formed by baryonic outflows, and in which density profiles remain steep  in the lowest mass galaxies. Interestingly, the low velocity end of  observations also show an upturn, as derived in \citet{klypin14} and precisely as predicted by the DC14 model.

Of course, the upturn in the observations needs to be taken with a grain of salt. 
Indeed,  moving to lower masses, observations are less well constrained for two reasons. Firstly, below stellar mass of $\sim$10$^9$\msun\ the HI disks become on average thicker \citep{brinks02} with the velocity dispersions being a larger fraction of the total rotation velocity,  making the interpretation of $W50$ increasingly difficult and the approximation of where to measure V$_{\rm rot}$ in the models less well defined. Secondly, at these masses a correction for completeness is required to account for the increasingly significant number of galaxies that have no measured HI \citep{klypin14}. Future surveys with larger samples, more accurate velocity measurements, detailed accounting for completeness  and precise understanding of the radius at which velocities are measured, are required in order to determine whether there is a statistically solid    upturn in the \vlos\ distribution at \vlos$<$20$\,$km$\,$s$^{-1}$.

\begin{figure*}
\hspace{0.cm}\includegraphics[width=.8\textwidth]{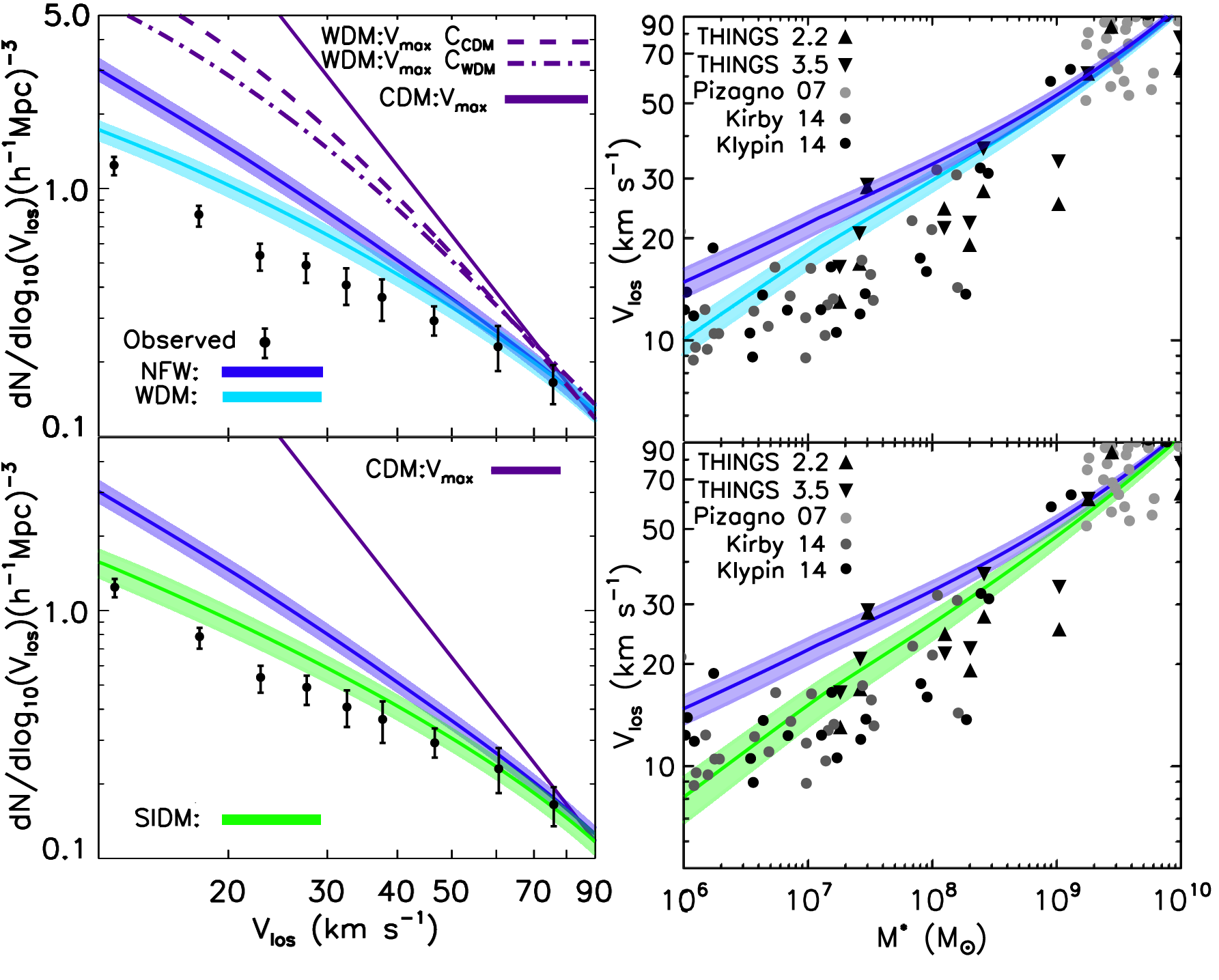}
\caption{Top Left Panel: The velocity function, or \vlos\ distribution, in a 2$\,$KeV WDM model shown as light blue region, with velocities measured at the radius indicated in Eq.~\ref{rm}. In the case of velocities measured at \vmax\ the resulting \vlos\ distribution is shown as dashed purple line (concentraction-mass relation from CDM) or dot-dashed purple line (concentration corrected for WDM, as in \citet{schneider12}). The main effect of WDM particles is to change the halo mass function with respect to the CDM case. Top Right Panel: The \mstar-\vlos\ relation for WDM, with \vlos\ measured between 2.2 and 3.5 disc scale-length (shaded region, with solid line representing the 2.85 scale-length measurement). Bottom Left Panel: The \vlos\ distribution for cored profiles representative of the adopted SIDM models, as green area. Core sizes increase  in proportion to the decrease in disc scale length from zero at \mstar=6$\times$10$^{10}$\msun\ to a maximum core size of 1$\,$kpc at \mstar$=$10$^8$\msun. Bottom Right Panel:  The \mstar-\vlos\ relation for SIDM.
Alternative DM models result in  a flattening of the \vlos\ distribution and  a downward bend in the \vlos-\mstar\ relation, compared to the \lcdm\ model with NFW profile,  but are unable to explain observations over the full mass range.}
  \label{fig:altDM}
\end{figure*}

Bearing these caveats in mind,  the comparison between the theoretical  and  observed velocity function highlights three points:  \\
\noindent (i) the cold dark matter model where haloes follow the widely used NFW density profile is ruled out: even by including a correction for the radius at which velocities are measured, the data could not be reconciled with a universal NFW model,\\
\noindent (ii) the discrepancy between observations and theory can be easily explained as a manifestation of the cusp/core problem, with the mass dependent DC14 profile capturing the essence of it: a notable agreement between theory and observations is found if such a model is applied, and\\
\noindent (iii) the dark matter cores in galaxies formed by energetic baryonic outflows predict an upturn in the velocity function at low velocities, corresponding to the low mass galaxies regime where the energy form supernovae is less efficient at expanding haloes: such a feature may be tested with a careful comparison between theory and future updated observations.

 We note that the results at the low mass end are not particularly sensitive to our disc scale-lengths, as stars become a decreasing fraction of total mass. At the high mass end, the position where the the model \vlos\ intersects with the dark matter only \vmax, is sensitive to the adopted disc scale-lengths, as well as to the assumed adiabatic contraction,  but this does not affect our main results.

\subsection{\mstar-Velocity Relation}
In this section, we take another perspective of the problem, exploring the relation between the velocity and stellar mass of galaxies \citep{tully77}. Observational and systematic biases in this process will differ from  biases that may exist in the velocity distribution comparison made above. 

Galaxy stellar masses  from large surveys \citep{baldry12} can be matched to dark matter haloes \citep{guo10}, allowing us to derive theoretical rotation velocity for galaxies of given stellar mass at any radius.
We then obtain high resolution velocity information for a sample of observed galaxies \citep{pizagno07,kirby14,klypin14,oh08,walter08,deblok08,oh11} and compare observed and theoretical values of \vlos\  between 2.2 and 3.5 scale lengths. To derive this region, we use the empirical  relation between disc scale-lengths and \mstar\ \citep{dutton11b,fathi10}. 

In the right  panel of  Figure~\ref{fig:CDM} we show results of this comparison.  THINGS galaxies are plotted at both 2.2 (upward pointing triangles) and 3.5 scale-lengths (downward pointing triangles) \citep{walter08,oh08,deblok08,oh11}. At the high mass galaxies end \citep{pizagno07}, there is very little difference between measuring \vlos\ at 2.2 and 3.5 scale-lengths. At the low mass end, the region of 2.2-3.5 disc scale-lengths is  a reasonable proxy for the radial range at which kinematics of dwarf spheroidal galaxies are measured \citep{kirby14}.

As shown in the right  panel of Figure~\ref{fig:CDM}, the \lcdm\ model with NFW profile (blue shaded line) predicts that dwarf galaxies with \mstar $\sim$5$\times$10$^6$$-$5$\times$10$^8\,$M$_\odot$, should have significantly higher \vlos\ than  is observed, a well documented issue within the standard \lcdm\ model \citep[e.g.][]{read06,oh11,tollerud14}. 
The DC14 model (red shaded line), instead, predicts that the dark matter density profile is flattest for massive dwarf galaxies, such that the relation between \vlos-\mstar\ becomes lower than the NFW expectation at masses $10^7$$<$\mstar/M$_\odot$$<$$10^9$.
Moving to even lower stellar masses, the steepening of the halo profile causes the \vlos-\mstar\ relation to  flatten,  eventually re-joining the NFW model for galaxies with stellar masses of  \mstar$\sim$10$^6$\msun.
                   
The well known independence of velocity from stellar mass observed in low mass dwarf galaxies, such as satellites of the Milky Way, Andromeda and ultrafaint dwarfs (\citealt{strigari08,gk14}, and references therein), is therefore a feature naturally reproduced by haloes that are flattened by astrophysical processes, such as supernovae-energy-driven outflows of gas, and that retain a cuspy profile at the smallest scales, \mstar$<3\times10^6$\msun\ \citep{governato12,brooks14,DiCintio2014a}.
The overall result is that the DC14 model can explain the observed \vlos\ of dwarf galaxies over a wide range in stellar mass.

 A significant caveat must be mentioned here, regarding the observed measures of galaxy velocities. For the dwarf spheroidal galaxies,   the circular velocity has been approximated using V$(r_{1/2})=\sqrt{3\langle\sigma^2\rangle}$, which  provides a reasonable estimate at the half light radius of non-rotating, dispersion supported galaxies, minimizing the errors introduced by uncertainties of the anisotropy parameter \citep{walker09,wolf10}. This methodology has been shown to work also in non-spherically symmetric systems \citep{thomas11}. 
Systematic differences exist between these estimates of circular velocity, and those coming from rotation curve data. 
We therefore do not want to overstate any comparison between the predictions and observations. The purpose of the paper is to highlight ways that competing theories may be contrasted: it remains observationally challenging to test for the differences that we highlight, over the full mass range that may be required.  

\subsection{Alternative Dark Matter Models }

 Alternative  dark matter particles have also been suggested as  solutions to the velocity function problem \citep{zavala09,vogelsberger12,schneider14,klypin14,papastergis14}.
Such models  make different predictions for how halo density profiles vary with galaxy mass. 

Here,  we explore predictions for  the warm dark matter (WDM) and self interacting dark matter (SIDM) models introduced in Section~{2}. 
In the top left panel of  Figure~\ref{fig:altDM} we plot the velocity function derived assuming a 2$\,$KeV WDM particle, which results in a different halo mass function when compared to CDM predictions.
We show this effect measuring velocities at \vmax\ and using a CDM concentration-mass relation (dashed purple line).
We then make adjustments for WDM concentrations \citep{schneider12} including adiabatic contraction (dot-dashed purple line). 
Clearly, modification to the halo mass function driven by a ~2$\,$KeV particle is not enough to reconcile theory and observations, if haloes' velocities are measured at \vmax.

Therefore we procede to analyze a WDM model which includes the universal dark matter fraction plus a stellar and a gas discs, as presented in Section~{2}, and in which halo velocities are measured according to Equation~\ref{rm}.
We show the resulting velocity function in light blue in the top left panel of Figure~\ref{fig:altDM}. 
In the top right panel we show the \mstar-\vlos\ relation for WDM where the \vlos\ has been measured between 2.2 and 3.5 disc scale-lengths.

The bottom left panel  shows the \vlos\ distribution for cored profiles (green) representative of the adopted SIDM models \citep{vogelsberger12}. Core sizes in the model increase  in proportion to the decrease in disc scale length from zero at \mstar=6$\times$10$^{10}$\msun\ to a maximum core size of 1$\,$kpc at \mstar$=$10$^8$\msun.    The \mstar-\vlos\ relation for the SIDM model is shown in the bottom right panel.

\begin{figure}
\hspace{-.6cm}\includegraphics[width=.53\textwidth]{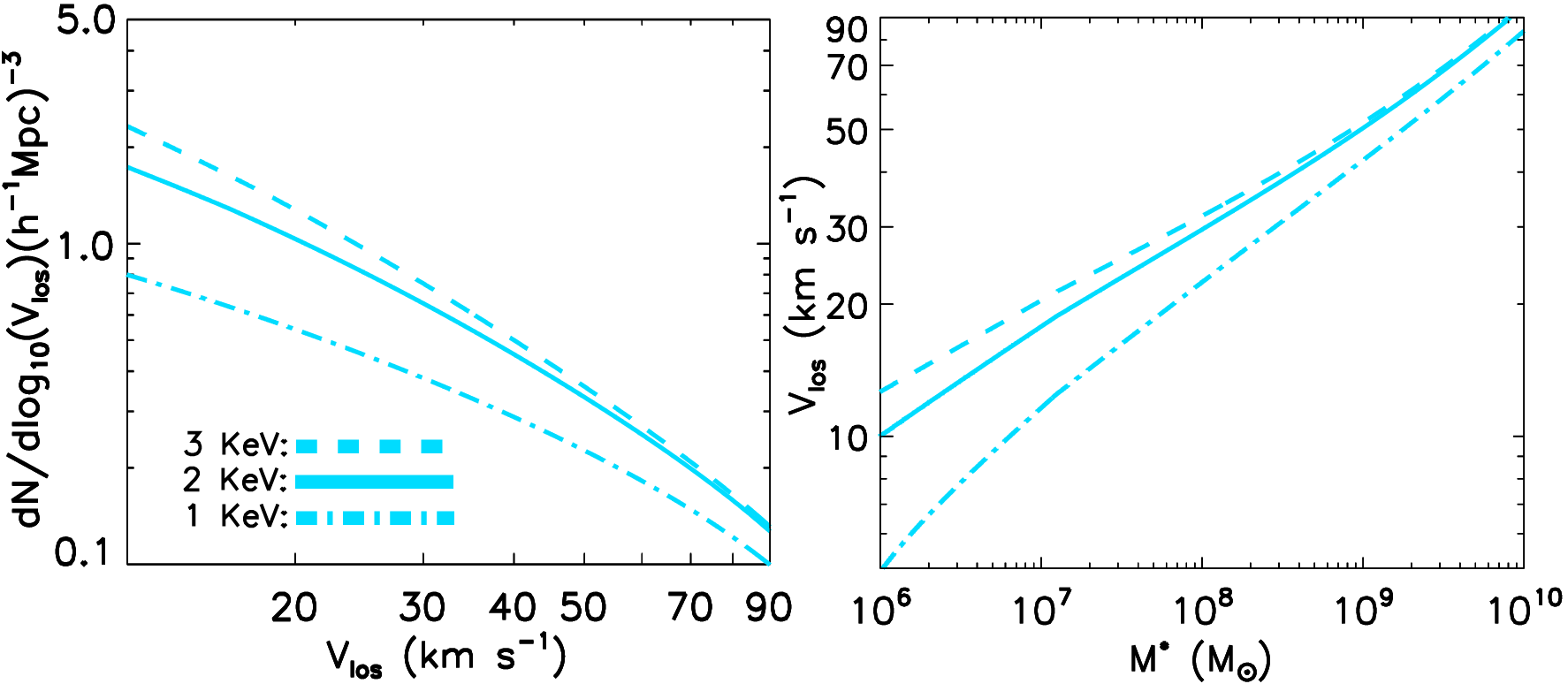}
\caption{Left Panel: The velocity function  for WDM particles of 1,2 \& 3 KeV is shown as dot dashed, solid and dashed line respectively. Right Panel: The \vlosMs\ relation for the different WDM particle masses.  The trends are the same in all cases, i.e. the \vlos\ distribution bends down compared to the \lcdm\ model as \vlos\ become smaller, while the \vlosMs\ relation is steeper than the \lcdm\ relation. 
 }\label{fig:wdm}
\end{figure}

Both these alternative dark matter theories result in a flattening \citep{zavala09} of  the   \vlos\ distribution  in the region 30$\lsim$\vlos$\lsim$60 km$\,$s$^{-1}$, as seen in  Figure~\ref{fig:altDM}.  As velocities become even  smaller, however,  the \vlos\ distribution  of these models increasingly diverge from the \lcdm\ distribution, contrary to what occurs  in the DC14 model. This finding is not specific to the WDM and SIDM models we show in Figure~\ref{fig:altDM}, but is  general to any alternative DM  model where core size increases relative to the virial radius, are constant at all masses, or scale with the virial radius of galaxies. 
 No such model is able to match observations over the entire velocity range.

Alternative DM models that predict a  decrease in the number of low mass haloes, and/or flattened halo density profiles even in the lowest mass galaxies, result in the velocity-\mstar\ relation bending down compared to the standard \lcdm\ case,  better matching the velocities of observed high mass dwarf galaxies (right panels in Figure~\ref{fig:altDM}). However, as we move to even lower mass,  such models continue to bend downward and do not match the velocity values observed for  dwarf galaxies with low stellar mass. The better such models are able to account for the velocity of high mass dwarfs, the worse they fare with respect to low mass dwarfs.  Our results can be generalized to rule out any model where core size increases monotonically as halo mass decreases.

\subsubsection{Different WDM particle masses}\label{wdmvary}
We show the \vlos\ distribution and \vlosMs\ relations  for WDM particles of 1, 2 \& 3 KeV as dashed, solid and dot dashed lines in Figure~\ref{fig:wdm}. The trends are the same in all cases, i.e. the \vlos\ distribution bends down compared to the \lcdm\ model as \vlos\ become smaller, while the \vlosMs\ relation is steeper than the \lcdm\ relation. None of the particle masses considered is able to reproduce observations, consistently with what has been shown previously \citep{klypin14,schneider14}. Moreover, we note that constraints from the Lyman-alpha forest indicate that WDM particles should have mass $\gsim$3$\,$KeV \citep{viel13}.

\begin{figure}
\hspace{-.6cm}\includegraphics[width=.53\textwidth]{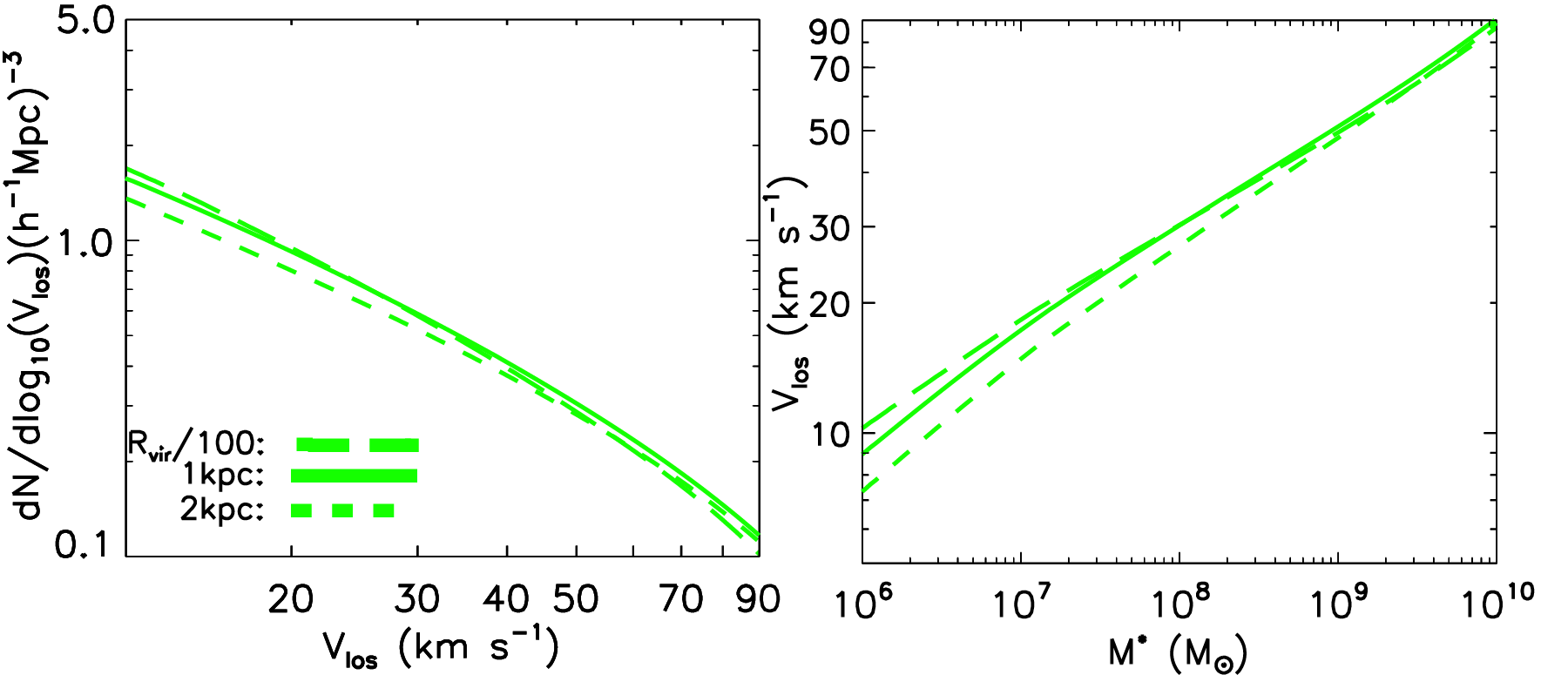}
\caption{ The \vlos\ distribution (left panel) and   \vlosMs\ relation (right panel) for various core sizes and scalings with mass. Core sizes increase linearly in proportion to the decrease in disc scale length from zero core at \mstar=6$\times$10$^{10}$\msun\ to a maximum core size of 1 (solid green line) and 2 (dashed green line)$\,$kpc at \mstar$=$10$^8$\msun. The purple dashed line shows cores which scale in size with virial radius, core size = R$_{\rm vir}$/100. 
 }
 \label{fig:cores}
\end{figure}

\subsubsection{Different core sizes}\label{corevary}
The parameterization of core sizes with  mass, used within the adopted SIDM model in Section 3.3,  mimics the main features of  core dependence on halo mass for velocity dependent SIDM particles \citep{vogelsberger12}. In Figure \ref{fig:cores} we show the results for different parameterizations of core sizes with mass.  Core sizes increase linearly in proportion to the decrease in disc scale length from zero core at \mstar=6$\times$10$^{10}$\msun\ to a maximum core size of 1$\,$kpc (solid green line) and 2$\,$kpc (dashed green line) at \mstar$=$10$^8$\msun.

 Keeping a constant 1 or 2$\,$kpc core at high masses makes very little difference (making the fit to observations at the high mass end slightly worse), so is not shown. We also show a core which scales in size with virial radius, such that core size = R$_{\rm vir}$/100, which results in less flattening of the \vlos\ distribution. None of the used parameterizations is able to reproduce the \vlos\ distribution and the  \vlosMs\ relation over the entire mass ranges.

\subsection{AGN and  cores in massive disc galaxies.}
The large amounts of energy from AGN in the inner region
of galaxies could lead to a further transformation of the inner dark matter profile \citep{martizzi13}. 
We did not include AGN in our study.
Firstly, AGN activity is not likely
to be significant in the mass range that is relevant to this study.

Secondly, baryons make an increasingly significant contribution to the inner
density profiles at high masses: the differences in the dark matter
density profile are therefore less important in high mass galaxies.
We verified that a model which follows the mass dependence
of the DC14 density profile at low mass, but has cored profiles for
all galaxies with \mstar$>$3$\times$10$^8$\msun, in order to resemble
a possible AGN-driven core, does not change our results.

\subsection{Effects of Abundance Matching Uncertainties}\label{section:ab}
\begin{figure}
\hspace{.0cm}\includegraphics[width=.45\textwidth]{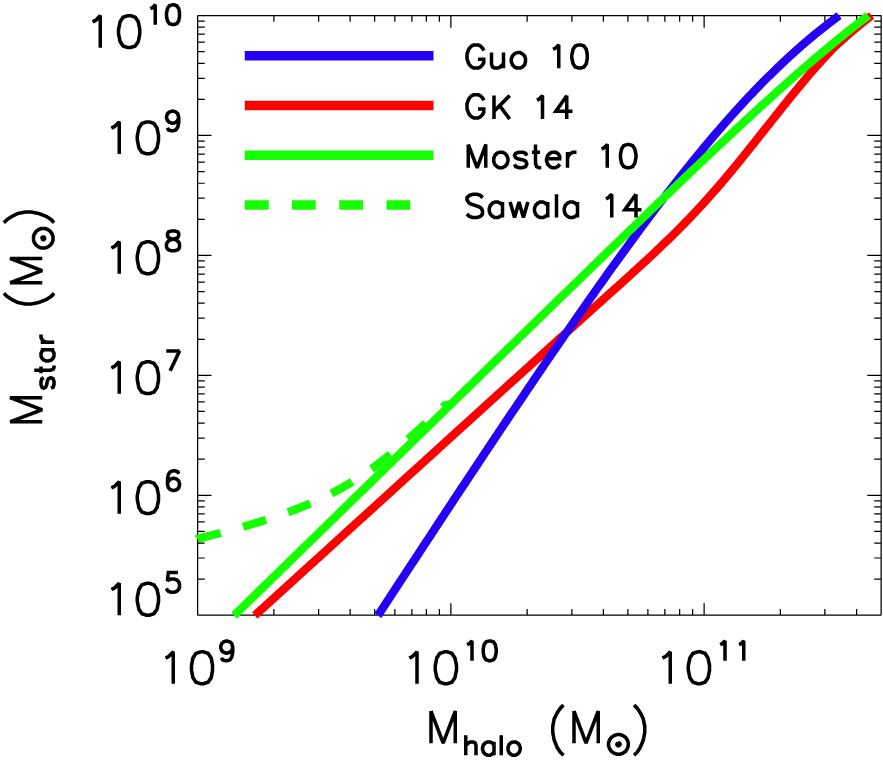}
\caption{The \mstar\ versus \mhalo\ relations \citep{guo10,gk14,moster10,sawala14} that are explored. These relations cover a wide range of normalisations and slopes in the region relevant for our study. }
\label{fig:AM}
\end{figure}
In this study we relate stellar masses and halo masses using an empirical abundance matching relation \citep{guo10}. Several other studies provide different formulations for such relation. Here we show that the particular form of the abundance matching relation that we employ is not driving our results and conclusions.

Abundance matching relations  are generally complete down to a stellar mass of few \mstar$\sim10^8$\msun, corresponding to the lower limit of  large scale surveys such as SDSS \citep{baldry08} and GAMA \citep{baldry12}. Above this mass, there is relatively little difference in the abundance matching studies \citep{guo10,moster10,behroozi13}, and such differences are insignificant in terms of our study. 
However for stellar masses below $\sim$10$^8$\msun\ observational uncertainties exists and, as this is a crucial  mass range in our study, we checked if using different abundance matching prescriptions affects the results at  low stellar masses.

There are two recent studies that have extended the abundance matching relation to masses lower than \mstar$\sim10^8$\msun, by using the observed stellar mass function of the Local Group \citep{brook14,gk14}. In the first study, \citet{brook14}, it was shown that using the average halo mass function of simulated local groups, which is well described by a power law, implies  a steep relation between \mstar-\mhalo\ in the region 10$^{6.5}$$<$\mstar/\msun$<$10$^8$. The empirically  extended  relation found in \citet{brook14} matches well the extrapolated abundance matching relation used in this paper \citep{guo10}.

The second study, \citet{gk14}, choses a particular collisionless simulation of the local group which has a flatter than average halo mass function,   allowing the use of a  flatter \mstar-\mhalo\ relation when  matching haloes to  the Local Group stellar mass function. It also showed that,  using an even  flatter \mstar-\mhalo\  relation as in \citet{behroozi13}, the results is a severe over-estimate of the number of local group galaxies with stellar masses 10$^{6.5}$$\lsim$\mstar/\msun$\lsim$10$^8$.

 Considering these empirical constraints, we explore the effects of choosing flatter abundance matching relations than the one used in the paper. 
 The three tested relations \citep{guo10,gk14,moster10}, shown in Figure~\ref{fig:AM},  cover a very broad range in slope and normalisation, particularly at low masses. For a halo mass of 10$^{10}$\msun, there is approximately an order of magnitude difference in the matched stellar mass according to the different relations used.
The velocity function relation of the DC14, NFW, SIDM and WDM models will only suffer from a negligible change when different abundance matching  are applied, and therefore it is not shown here.
\begin{figure}
\hspace{-.1cm}\includegraphics[width=.47\textwidth]{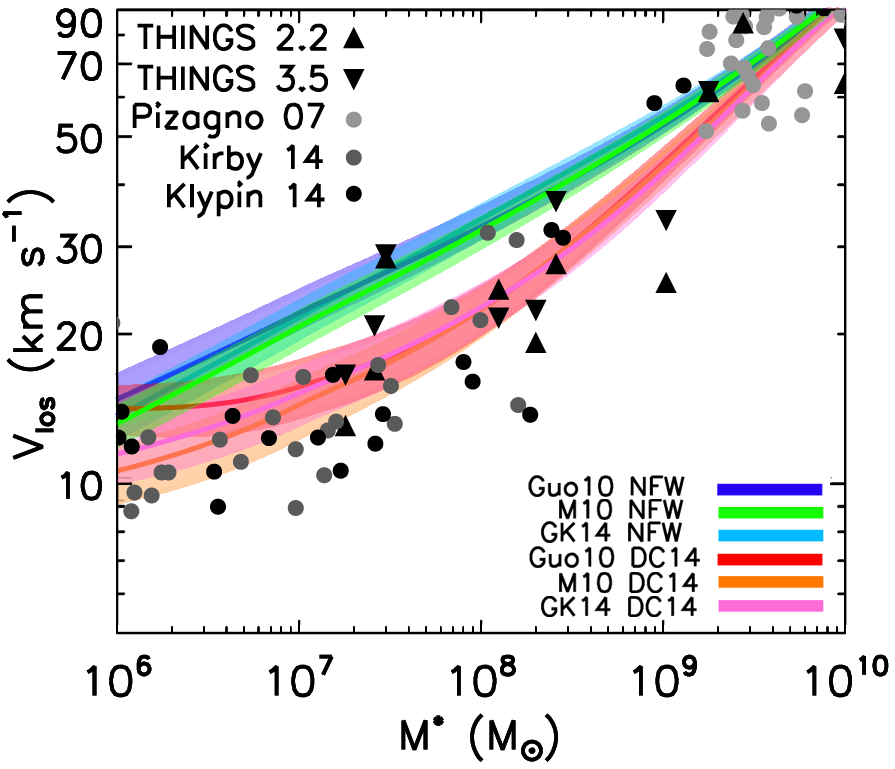}
\caption{The \mstar-\vlos\  relations for the  NFW and DC14 models,  using different abundance matching relations \citep{guo10,gk14,moster10}, which span a wide range of normalisation and slope. The conclusions of the paper are not affected: the DC14 profile still provides a very good fit to the data, for all the abundance matching relations, while the NFW profile does not match observations when the full range of  dwarfs masses is considered.}
\label{fig:AMrel}
\end{figure}
We plot  \mstar-\vlos\ using  the different abundance matching relations for the NFW and DC14 models in Figure~\ref{fig:AMrel}, and for the alternative dark matter models in Figure~\ref{fig:S9}. 

The most important general observation is that altering the abundance matching will alter each of the models in a consistent manner. To be specific, choosing an abundance matching relation that is flatter than the fiducial one used in the paper \citep{guo10},  steepens the \mstar-\vlos relation in all models. Thus, all models that were already too steep  in the region  5$\times$10$^6$$<$\mstar/\msun$<$3$\times$10$^8$, (i.e. all but the DC14 model),  worsen. The DC14 model still does very well at matching the observations in this important mass range,  and the advantage over the other models is clearly maintained in full.  The  NFW, SIDM, and WDM models all fail at reproducing the observed \mstar-\vlos\ relation, predicting dwarf galaxies with an increasingly low \vlos\ when a flatter abundance matching relation is assumed.

It may seem a  little surprising that a large change in the abundance matching relation results in a relatively small change in the \mstar-\vlos\ relation. This is in part due to lower mass haloes having higher concentrations. We remind that the fact that the mass-concentration relation results in similar velocity measurements, for low mass galaxies  over a range of halo masses, has been shown previously \citep{kravtsov10}.
This study concerns the effects of baryons in altering the haloes in which galaxies are hosted, which we implement using the DC14 model. A different proposal is that  baryonic processes  effectively ``adjust" the halo mass function \citep{sawala14}, flattening the  \mstar-\mhalo\ relation. In the most important region of our study,  5$\times$10$^6$$<$\mstar/\msun$<$3$\times$10$^8$,  galaxy stellar masses are matched to haloes that are well above the mass where  re-ionization  may render some haloes dark, \mhalo$\sim$5$\times$10$^{9}$\msun\ \citep{bullock00},  regardless of which abundance matching relation is used. The ``bend" in the \mstar-\mhalo relation due to re-ionisation, as proposed by \cite{sawala14}, occurs at lower masses than what is considered in this study, \mstar$<$10$^6$\msun.

\section{Conclusions}
The  \textit{cusp-core} discrepancy is a challenge to the \LCDM\ paradigm, which is heightened by studies showing that non-standard dark matter particles can result in cored dark matter haloes. In this paper we have highlighted how it is possible, from a theoretical point of view, to distinguish between astrophysical processes and alternative dark matter scenarios that are able to create cores in galaxies.

The key is that different models make different predictions for how core formation depends on mass.
We show the signatures of these differences in galaxy populations, in particular by studying the velocity function and the Tully-Fisher relation.

Using analytic models, we have shown that the density profiles, DC14, resulting from energetic feedback processes like supernova explosions, have a particular mass dependence with a maximum core formation efficiency at \mstar$\sim$10$^{8.5}$\msun\ \citep{DiCintio2014a,DiCintio2014b}. This value matches the peak mass of core formation occurring in hydrodynamical simulations \citep{stinson13,brook12}. These findings allow us to create model populations of galaxies within dark matter haloes that have been affected by such energetic processes, and to compare them with models that assume a universal, steep, NFW density profile.

\begin{figure}
\hspace{-.6cm}\includegraphics[width=.55\textwidth]{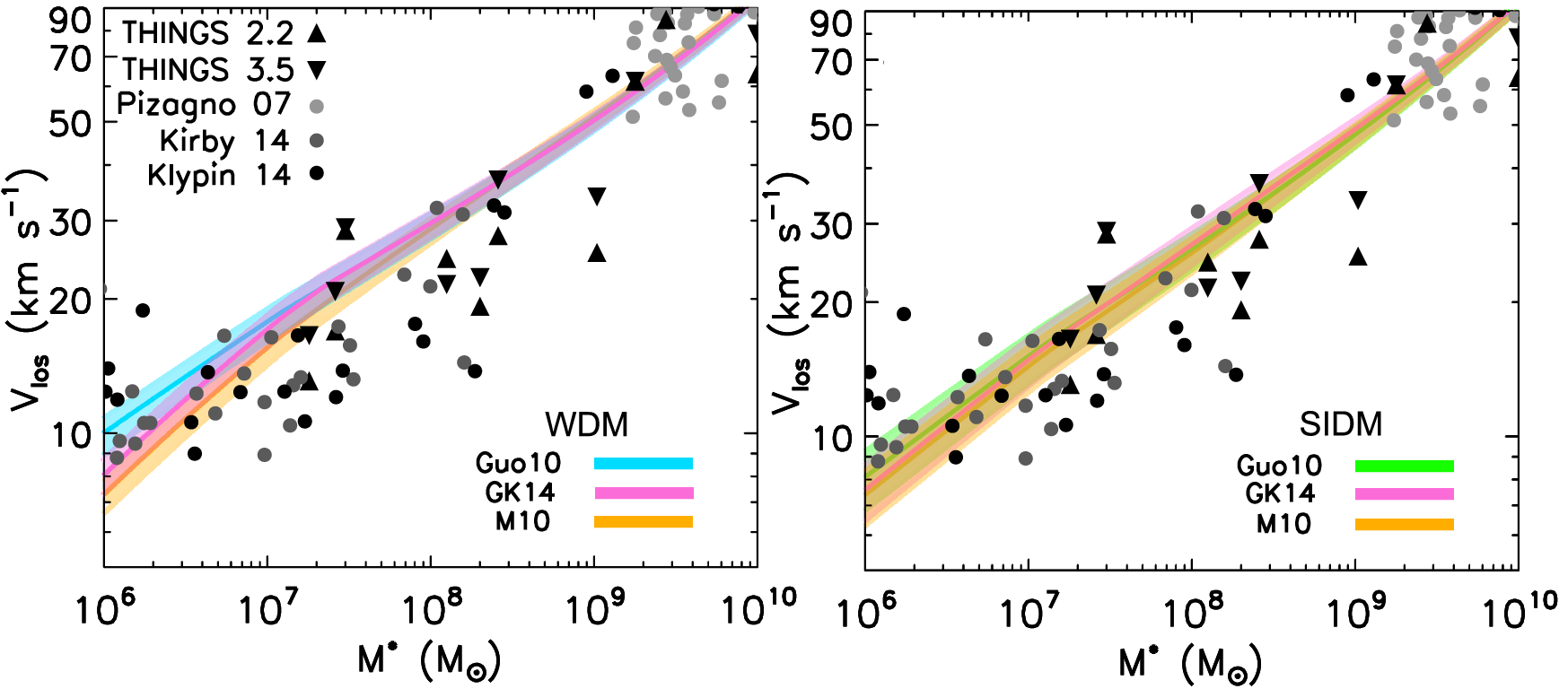}
\caption{The \mstar-\vlos\  relations for the  WDM and  and SIDM models,  using different abundance matching relations  \citep{guo10,gk14,moster10}. A flatter relation between  \mstar\ and \mhalo\ results in a steeper \mstar-\vlos\ relation, worsening the mismatch with observations. }\label{fig:S9}
\end{figure}

In agreement with previous studies, we showed that it is difficult to reconcile a model galaxy population that assumes an NFW profile for dark matter haloes with the observed  velocity function of galaxies with line-of-sight velocity \vlos$<$60 km/s \citep{zavala09,papastergis11,tg11,klypin14,papastergis14}, and with the observed Tully-Fisher relation of galaxies with stellar mass $\sim$10$^7$-10$^9$\msun. The mismatch holds without assuming a priori that galaxies rotation curves are tracking the maximum velocity of dark matter haloes.

By contrast, a model in which the density profile varies according to the ratio \mstar/\mhalo, as expected when the halo profile is affected by baryonic outflows of gas,  is able to reproduce both the velocity function and the \mstar-velocity relation of galaxies, under plausible assumptions regarding the manner in which observations and theory are compared.
Further, we show that the particular mass dependence of cores formed by baryonic outflows will leave signatures within both the velocity function and the \mstar-velocity  relation of galaxies. Specifically, the steepening of the profiles as galaxy mass decreases below \mstar$\sim$10$^{8.5}$\msun\  results in an upturn in the velocity function for \vlos$\lsim$20$\,$\kms\ and in a flattening of the \mstar-velocity relation in the region 10$^{6}$$\lsim$\mstar/\msun$\lsim$10$^{8}$, both features in agreement with current data \citep{klypin14}.

The  independence of velocity from stellar mass observed in low mass dwarf galaxies (\citealt{strigari08,gk14}, and references therein) is thus a feature naturally reproduced by haloes that are flattened by astrophysical processes.
Evidence for an observed upturn in the velocity function is more tenuous,  relying on the manner in which galaxies that are devoid of gas are added to results coming from large scale surveys of HI line-widths.  Further to this, an increased ratio of velocity dispersion to rotation velocity means that HI line-widths of low mass galaxies may be more difficult to interpret than in more massive galaxies.

Alternative dark matter models also make  predictions for how halo density profiles vary with galaxy mass.  We explored predictions from WDM and SIDM models, finding that both result in a flattening of  the   \vlos\ distribution in the region 30$\lsim$\vlos$\lsim$60 km$\,$s$^{-1}$;  as velocities become even  smaller, however,  the \vlos\ distribution  of these models increasingly diverge from the \lcdm\ distribution, contrary to what occurs  in the DC14 model. 
Any model that predicts a  decrease in the number of low mass haloes (such as WDM) and/or cores in the lowest mass galaxies (such as SIDM), results in the \mstar-velocity relation better matching observations of high mass dwarf galaxies than the NFW model, but as we move to the lowest  masses, such models continue to bend downward and do not match the observed velocities of the faintest dwarf galaxies. The better such models are able to account for the velocity of high mass dwarfs, the worse they fare with respect to low mass dwarfs.  

These findings can be generalized to any alternative dark matter model in which core size increases monotonically as halo mass decreases,  are constant at all masses, or scale with the virial radius of galaxies.

It must be noted that in this paper we did not use the distribution of gas and stars from hydrodynamical simulations, but  we rather coupled  the mass dependent DC14 profile with empirical galaxy scaling relations to make model galaxy populations. 

In a future study we will compare HI line-widths measured directly from our simulations to the empirical $\rm V_{max}-V_{rot}$ abundance matching relation found in  \citet{papastergis14}. We note that the hydrodynamical simulations of \citet{governato12} and \citet{brooks14} included in \citet{papastergis14} did not match the  $\rm V_{max}-V_{rot}$ abundance matching at the low mass end, and we will explore whether this holds in our simulations and whether there are differences between simulated galaxies and  empirical model for galaxy properties adopted here.

  It remains a theoretical and observational challenge to compare measured  velocities of observed dwarf galaxies to model predictions, particularly as the ratio of  velocity dispersion to rotational velocity increases for low mass galaxies.
 Uncertainties associated with stellar velocity dispersions in local dSphs should be reduced thanks to GAIA satellite data \citep{perryman01}, and  possibly also  next generation 40m telescopes such as the European ELT, working with multi adaptive optic cameras \citep{davies10}.
 Meanwhile, the next generation of HI surveys, as we enter the era of SKA,  promise high  spatial resolution  HI gas kinematics in a large sample of galaxies, with increased fidelity in the derivation of the velocity function \citep{lister15}.

 More sophisticated comparisons between observations and theory are certainly required, as are further models exploring the nature of density profiles that form under the influence of baryonic outflows; different feedback implementations, higher resolution simulations and more diverse initial conditions will likely result in an improvement of our understanding of how density profiles are modified by baryonic physics.

\section*{Acknowledgements}
We thank Manolis Papastergis and Simon White for very helpful comments on an earlier draft. 
CB thanks the MICINN (Spain) for the financial support through the MINECO grant AYA2012-31101 and  the Ramon y Cajal program. He further thanks the DARK cosmology centre for the kind hospitality. ADC is supported by the DARK independent fellowship program.

\bibliographystyle{mn2e}
\bibliography{archive}


\label{lastpage}

\end{document}